\documentclass[a4paper]{jpconf}
\usepackage{graphicx}
\usepackage{times,amsmath,amsfonts,amssymb,bbm,stmaryrd,simplewick,mathrsfs,mathtools,dsfont,slashed,wrapfig}
\usepackage{hyperref}

\newcommand{\C}{\mathds C}
\newcommand{\N}{\mathds N}
\newcommand{\RR}{\mathds R}

\newcommand{\cG}{\mathcal{G}}

\def\im{\mathrm{i}}
\def\ep{\mathrm{e}}
\def\pa{\partial}
\def\diff{\mathrm{d}}

\def\tr{\mathrm{tr}}
\def\sfrac#1#2{{\textstyle\frac{#1}{#2}}}
\def\>{\rangle}
\def\<{\langle}
\def\+{\dagger}

\def\={\ =\ }
\def\unity{\mathbbm{1}}
\def\und{\quad\textrm{and}\quad}
\def\with{\quad\textrm{with}\quad}

\begin{document}

\title{The Nicolai-map approach to supersymmetry}

\author{Olaf Lechtenfeld}

\address{Institut f\"ur Theoretische Physik, Leibniz Universit\"at Hannover, Appelstrasse 2, 30167 Hannover, Germany}

\ead{olaf.lechtenfeld@itp.uni-hannover.de}

\begin{abstract}
In 1980 Hermann Nicolai proposed a characterization of supersymmetric theories that
became known as the Nicolai map. This is a particular nonlocal and nonlinear field
transformation, whose perturbative expansion is given by fermion-line trees with bosonic
leaves. Quantum correlation functions can by evaluated using the inversely transformed
fields in the free theory. After initial promise and excitement (fuelling the author's PhD
work!), the subject all but fell dormant for 35 years. Recently however, technical progress
in the construction as well as a deeper insight into the nature of the map have been
achieved, from quantum mechanics to super Yang--Mills in various dimensions. I will present 
the Nicolai map from this modern perspective and touch on some of the current developments.
\end{abstract}

\section{A question raised in 1980 (by Nicolai) and answered until 1984 (but not fully ...)}

The key idea is best illustrated by an example.
Let us look at the Wess--Zumino model in $3{+}1$ dimensional Minkowski space,
consisting of a complex scalar~$\phi$, a Weyl fermion~$\psi$ and a complex auxiliary~$F$,
characterized by a superpotential~$W(\phi)$ and defined by the off-shell lagrangian~\footnote{
A multi-field generalization is straightforward.}
\begin{equation}
{\cal L} \= \pa_\mu\phi^*\pa^\mu\phi + F^*F + 
\tfrac{\im}{2} \bar\psi\bar\sigma{\cdot}\pa\psi - \tfrac{\im}{2} \psi\sigma{\cdot}\pa\bar\psi +
W'(\phi)\,F + W'(\phi)^*F^* - \tfrac12\psi\,W''(\phi)\,\psi - \tfrac12 \bar\psi\,W''(\phi)^*\bar\psi\ ,
\end{equation}
where $(\sigma^\mu)=(\unity,\vec{\sigma})$ and $(\bar\sigma^\mu)=(\unity,-\vec{\sigma})$
with Pauli matrices~$\vec\sigma$.
Integrating out the auxiliary fields yields $F^*=-W'(\phi)$ and
\begin{equation}
{\cal L}_{\textrm{SUSY}} \= \bigl|\pa\phi\bigr|^2 -\bigl|W'(\phi)\bigr|^2 + 
\bigl( \tfrac{\im}{2} \bar\psi\,\bar\sigma{\cdot}\pa\psi - \tfrac12\psi\,W''(\phi)\,\psi + \textrm{h.c.}\bigr)\ .
\end{equation}
Integrating out the fermions $(\psi,\bar\psi)$ produces a functional determinant
$\det M=\exp\{\tfrac{\im}{\hbar}{\cdot}(-\im\hbar\,\tr\ln M)\}$
so that the action becomes
\begin{equation}
S_g[\phi] \= \smallint\!\diff^4\!x\ \bigl\{ |\pa\phi|^2 - |W'|^2 \bigr\}
\ -\ \im\hbar\,\tr\ln \Bigl(\begin{smallmatrix} 
W'' & \im\,\sigma{\cdot}\pa \\[4pt]
-\im\,\bar\sigma{\cdot}\pa & {W''}^* 
\end{smallmatrix} \Bigr)
\ =:\ S_g^b[\phi] + \hbar\,S_g^f[\phi]\ .
\end{equation}
Here, $g$ denotes some coupling constant(s) or parameter(s) inside the superpotential~$W(\phi)$.
The objects of desire are quantum correlators
\begin{equation} \label{Ycorr}
\bigl\< Y[\phi] \bigr\>_g \= \int\!\!{\cal D}\phi\ \ep^{\frac{\im}{\hbar} S_g[\phi]}\ Y[\phi]
\quad\with\quad \bigl\<\unity\bigr\> \= 1
\end{equation}
for any bosonic (local or nonlocal) functional~$Y$.

The path integral in (\ref{Ycorr}) describes a purely bosonic nonlocal field theory.
What is characteristic of its supersymmetric origin? In other words:
given such a nonlocal action~$S_g$, how could one infer its hidden supersymmetric roots?
This question was answered in 1980 by Hermann Nicolai~\cite{Nic1,Nic2,Nic3}:
Such hiddenly supersymmetric theories admit a nonlocal and nonlinear invertible map
\begin{equation} \label{Nicdef}
T_g:\ \phi\,\mapsto\ \phi'[\phi;g]
\qquad\textrm{such that}\qquad
\bigl\<Y[\phi]\bigr\>_g \= \bigl\<Y[T_g^{-1}\phi]\bigr\>_0 \quad \forall\,Y\ ,
\end{equation}
relating correlators in the interacting theory ($g{\neq}0$) to (more complicated) correlators
in the free theory ($g{=}0$). For the path integrals, this is equivalent to
\begin{equation}
{\cal D}\phi\ \exp\bigl\{ \tfrac{\im}{\hbar} S_g[\phi] \bigr\} \=
{\cal D}(T_g\phi)\ \exp\bigl\{ \tfrac{\im}{\hbar} S_0[T_g\phi] \bigr\} \=
{\cal D}\phi\ \exp\bigl\{ \tfrac{\im}{\hbar} S_0[T_g\phi] + \tr\ln\tfrac{\delta T_g\phi}{\delta\phi} \bigr\}\ .
\end{equation}
Separating powers of $\hbar$ in the exponent, this splits into two properties,
\begin{subequations}
\begin{equation} \label{freeaction}
S_0^b[T_g\phi] \= S_g^b[\phi] 
\qquad\textrm{``free action condition''} \ ,
\end{equation}
\begin{equation} \label{detmatching}
{}\quad S_0^f - \im\,\tr\ln\tfrac{\delta T_g\phi}{\delta\phi} \= S_g^f[\phi]
\qquad\textrm{``determinant matching condition''} \ .
\end{equation}
\end{subequations}
Every Nicolai map has to fulfil these two conditions, which originally were taken as its definition.
The reason for the name of~(\ref{detmatching}) is that its exponentiation gives an equality
of the functional fermion determinant~$\det M$ with the Jacobian of the transformation
(the first term is a constant since $S_0^f$ does not depend on~$\phi$).
From now on we put $\hbar{=}1$.

The Nicolai map provides an alternative characterization of supersymmetry or, more colloquially,
``supersymmetry without fermions''. Here is a sketch of its early history:
\begin{itemize}
\item[1979] the map for scalar theories, existence proof (incomplete)
\item[1980] refinement of the proof but a gap remains (Golterman 1982)
\item[1980] extension to gauge theories, construction to $O(g^2)$ [to $O(g^3)$ in 2020, $O(g^4)$ in 2021!]
\item[1982] examples of {\it linear\/} maps in $D{\le}2$, zero-mode obstruction for super Yang--Mills $\leftarrow$ Witten index)
\item[1984] linear maps for $D=4\ \&\ 6$ super Yang--Mills, doubtful due to ``Euclidean light-cone gauge''
\item[1983] $\!\!\!$/84\ \ constructive (perturbative) proof via coupling flow, requires off-shell supersymmetry
\end{itemize}

In 1984, the author derived (for his dissertation) an infinitesimal version~\cite{L1,DL1,DL2,L2}
of the Nicolai map by considering the
$g$-derivative of~(\ref{Nicdef}),
\begin{equation} \label{Nicinf}
\begin{aligned}
\pa_g\bigl\<Y[\phi]\bigr\>_g
&\ \stackrel{(\ref{Nicdef})}{=}\ \pa_g \bigl\<Y[T_g^{-1}\phi]\bigr\>_0 \\
&\=\, \bigl\<\pa_g Y[\phi]\bigr\>_g\ +\
\bigl\< \smallint (\pa_g T_g^{-1}\phi)\cdot\tfrac{\delta Y}{\delta\phi}[T_g^{-1}\phi]\bigr\>_0 \\
&\!\stackrel{(\ref{Nicdef})^{-1}}{=} \bigl\<\pa_g Y[\phi]\bigr\>_g\ +\
\bigl\< \smallint (\pa_g T_g^{-1}\circ T_g)\phi\cdot\tfrac{\delta Y}{\delta\phi}[\phi]\bigr\>_g
\ =:\ \bigl\< (\pa_g+R_g[\phi])\,Y[\phi] \bigr\>_g
\end{aligned}
\end{equation}
with a ``coupling flow operator''~\footnote{
We write $\diff x$ for the spacetime volume differential as long as its dimension remains unspecified.}
\begin{equation}
R_g[\phi] \= \int\!\diff x\ \bigl( \pa_g T_g^{-1}\circ T_g\bigr)\phi(x)\,\frac{\delta}{\delta\phi(x)}
\end{equation}
representing a functional differential operator derived from~$T_g$.

Nothing is gained, however, by these formal considerations, unless we can reverse the logic
and somehow obtain~$R_g$ and exponentiate it in order to create a finite coupling flow~$T_g$ from
$g'{=}0$ to $g'{=}g$, by inverting
\begin{equation}
\bigl(T_g^{-1}\phi\bigr)(x) \= \exp\bigl\{ g\,\bigl(\pa_{g'}+R_{g'}[\phi]\bigr)\bigr\}\,\phi(x)\,\big|_{g'=0}
\= {\textstyle\sum}_{n=0}^\infty\ \tfrac{g^n}{n!}\,\bigl( \pa_{g'} + R_{g'}[\phi] \bigr)^n\ \phi(x)\,\big|_{g'=0} \ .
\end{equation}
As we shall see in a moment, however, there exists a more direct construction of~$T_g$.
In any case, we have to establish the existence of the coupling flow operator~$R_g$
and find an explicit expression for it. We shall do this now for the exemplary case
of scalar theories (gauge theories will be treated in the following section).
If supersymmetry is realized off-shell on the action~$S$ then there exists a functional
$\mathring{\Delta}_\alpha[\phi,\psi,F]$
such that
\begin{equation}
\pa_g S[\phi,\psi,F] \= \delta_\alpha \mathring{\Delta}_\alpha[\phi,\psi,F]
\end{equation}
for the supersymmetry transformations~$\delta_\alpha$,
where $\alpha$ denotes a Majorana spinor index.
Integrating out the auxiliary~$F$ one has that
\begin{equation} \label{dgS}
\pa_g S_{\textrm{SUSY}}[\phi,\psi] \= \delta_\alpha \Delta_\alpha[\phi,\psi]
\qquad\textrm{with}\quad 
\Delta_\alpha[\phi,\psi] = \mathring{\Delta}_\alpha[\phi,\psi,-\smash{W'}^*(\phi)]
\end{equation}
for the on-shell action $S_{\textrm{SUSY}}\,{=}\int\!\!\diff x\,{\cal L}_{\textrm{SUSY}}$
with an anticommuting functional~$\Delta_\alpha$.
For our Wess--Zumino model example, it reads
$\Delta_\alpha=\tfrac12\int\!\diff^4\!x\,\psi_\alpha\,\pa_gW'(\phi)$.
The construction of~$R_g$ employs the supersymmetry Ward identity,
\begin{equation}
\pa_g \!\int\!\!{\cal D}\phi \!\int\!\!{\cal D}\psi\ \ep^{\im S_{\textrm{\tiny SUSY}}[\phi,\psi] }\ Y[\phi]
\= \int\!\!{\cal D}\phi \!\int\!\!{\cal D}\psi\ \ep^{\im S_{\textrm{\tiny SUSY}}[\phi,\psi] }\
\bigl( \pa_g + \im\Delta_\alpha[\phi,\psi]\ \delta_\alpha\bigr) Y[\phi]\ .
\end{equation}
Integrating out the fermions contracts bilinears to produce fermion propagators
$\bcontraction{}{\psi}{\,}{\psi} \psi\,\psi$ (in the $\phi$ background), hence~\cite{L1}
\begin{equation}
R_g[\phi] \= \im\,\bcontraction{}{\Delta}{_\alpha[\phi]\ }{\delta} \Delta_\alpha[\phi]\ \delta_\alpha \=
\im\int\!\diff x\ \bcontraction{}{\Delta}{_\alpha[\phi]\ }{\delta} \Delta_\alpha[\phi]\ \delta_\alpha \phi(x)\ 
\frac{\delta}{\delta\phi(x)}\ .
\end{equation}

\begin{figure}
\begin{center}
\includegraphics[width=0.7\paperwidth]{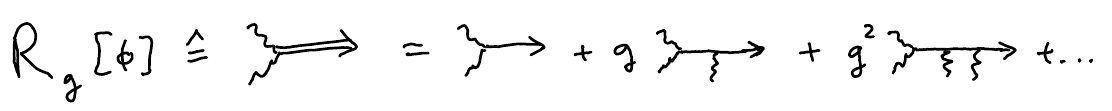} 
\vspace{-8mm}
\end{center}
\caption{\label{fig1}Graphical representation of the coupling flow operator}
\end{figure}

\begin{figure}
\begin{center}
\includegraphics[width=0.7\paperwidth]{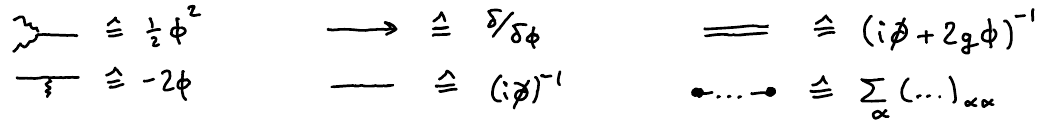}
\vspace{-7mm}
\end{center}
\caption{\label{fig2}``Nicolai rules'' for the diagrammatics}
\end{figure}

\begin{figure}
\begin{center}
\includegraphics[width=0.7\paperwidth]{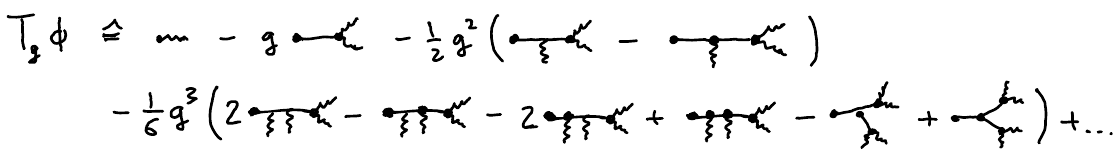}
\vspace{-6mm}
\end{center}
\caption{\label{fig3}Graphical representation of the Nicolai map to third order}
\end{figure}

\begin{figure}
\begin{center}
\includegraphics[width=0.7\paperwidth]{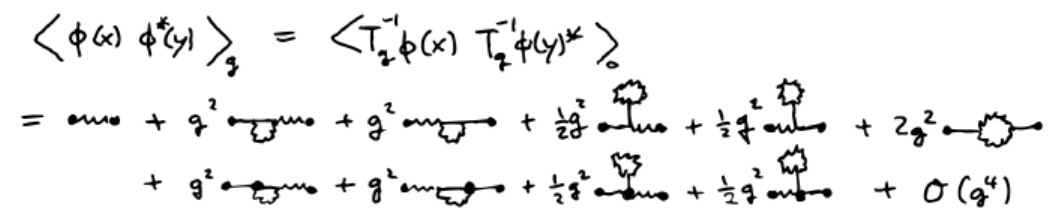}
\vspace{-6mm}
\end{center}
\caption{\label{fig4}``Nicolai diagrams'' for the two-point function}
\end{figure}

For a simple example of the Wess--Zumino model with (massless) superpotential $W=\tfrac13 g\phi^3$
in $3{+}1$ dimensions, one finds that
\begin{equation}
R_g[\phi] \= \tfrac{\im}{2} \smallint\!\!\smallint\diff^4\!x\,\diff^4\!y\ 
\bigl\{ \phi^2(x)\,\bcontraction{}{\psi}{(x)\ }{\psi} \psi(x)\ \psi(y) 
\ +\ {\smash{\phi^*}}^2(x)\,\bcontraction{}{\bar\psi}{(x)\ }{\psi} \bar\psi(x)\ \psi(y) \bigr\}_{\alpha\alpha}\,
\tfrac{\delta}{\delta\phi(y)} \ +\ \textrm{h.c.}
\end{equation}
where the subscript on the curly brace indicates a spin trace.
It is instructive to develop a diagrammatical shorthand notation.
For the sake of illustration, here we oversimplify $(\phi,\phi^*)\sim\phi$ and 
display the coupling flow operator as in Figure~\ref{fig1}
with graphical rules~\cite{FL} outlined in Figure~\ref{fig2}.
The linear tree for $R_g$ exponentiates to a series of branched trees for $T_g\phi$,
as exhibited in Figure~\ref{fig3},
and likewise for the inverse~$T_g^{-1}\phi$.
Inserting the latter into~(\ref{Nicdef}) and performing the free-theory bosonic contractions,
one obtains an alternative ``Nicolai'' perturbation series for correlators, as shown 
in Figure~\ref{fig4} for the two-point function.
Notably, the multiple action of~$R_g$ produces multiple spin traces
(graphically separated by dots).
The supersymmetric cancellation of the leading UV divergencies is automatically built in,
as pure fermion loops and boson tadpoles are absent by construction.

\section{A more universal answer in 2021}

Let us briefly focus on two important properties of the Nicolai map.
Firstly, $R_g$ is a derivation, and hence $T_g^{-1}$ acts distributively,
\begin{equation}
R_g[\phi]\,Y[\phi] \= \smallint \tfrac{\delta Y}{\delta\phi}\cdot R_g[\phi]\,\phi
\qquad\Leftrightarrow\qquad
T_g^{-1} Y[\phi] \= Y[T_g^{-1}\phi] \ .
\end{equation}
Secondly, by moving the map ``to the other side'',
\begin{equation}
\bigl\<Y[\phi]\bigr\>_0 \= \bigl\<Y[T_g\phi]\bigr\>_g\ ,
\end{equation}
choosing $\pa_g Y=0$ and differentiating with respect to~$g$, we learn that
\begin{equation}
0 \= \pa_g \bigl\<Y[T_g\phi]\bigr\>_g
\ \stackrel{(\ref{Nicinf})}{=}\ \bigl\<\bigl(\pa_g+R_g[\phi]\bigr)\,Y[T_g\phi] \bigr\>_g
\= \bigl\< \smallint \bigl(\pa_g+R_g[\phi]\bigr)\,T_g\phi \cdot \tfrac{\delta Y}{\delta\phi} [T_g\phi]\bigr\>_g
\end{equation}
for any (not explicitly $g$-dependent) functional~$Y$, and therefore necessarily
\begin{equation} \label{fixpoint}
\bigl(\pa_g + R_g[\phi]\bigr)\,T_g\phi(x) \= 0\ .
\end{equation}
This ``fixpoint property'' of the Nicolai map under the infinitesimal coupling flow allows us
to directly construct $T_g\phi$ from $R_g$ without invoking the inverse first~\cite{L2}.

Indeed, as had been missed in 1984, (\ref{fixpoint}) is formally solved by a path-ordered exponential,
\begin{equation} \label{universal}
T_g\phi \= {\cal P} \exp \Bigl\{-\!\!\int_0^g\!\!\diff h\ R_h[\phi]\Bigr\}\ \phi
\= \sum_{s=0}^\infty (-1)^s\!\!\int_0^g\!\!\diff h_s \ldots \!\int_0^{h_3}\!\!\!\!\diff h_2 \!\int_0^{h_2}\!\!\!\!\diff h_1\
R_{h_s}[\phi] \ldots R_{h_2}[\phi]\,R_{h_1}[\phi]\ \phi\ ,
\end{equation}
providing a ``universal formula'' for the Nicolai map in terms of the infinitesimal coupling flow~\cite{LR1}.
It is often useful to expand the flow operator in powers of the coupling,
\begin{equation}
R_g[\phi] \= \sum_{k=1}^\infty g^{k-1} r_k[\phi] \= r_1[\phi] + g\,r_2[\phi] + g^2 r_3[\phi] + \ldots
\end{equation}
from which one easily computes a power series expansion for the map itself,
\begin{equation} \label{Tseries}
T_g\phi \= \!\sum_{\bf n} g^n\,c_{\bf n}\,r_{n_s}[\phi]\ldots r_{n_2}[\phi]\,r_{n_1}[\phi]\ \phi
\qquad\!\textrm{with}\quad
{\bf n} = (n_1,n_2,\ldots,n_s)\ ,\quad n_i\in\N \ ,\quad {\textstyle\sum}_i n_i = n\ ,
\end{equation}
where $1\le s \le n$ and the $n{=}0$ term is the identity.
The numerical coefficients are computed as
\begin{equation}
c_{\bf n} \ = (-1)^s\!\!\int_0^1\!\!\diff x_s\;x_s^{n_s-1} \ldots
\!\!\int_0^{x_3}\!\!\!\!\diff x_2\;x_2^{n_2-1} \!\!\int_0^{x_2}\!\!\!\!\diff x_1\;x_1^{n_1-1} \=
\frac{(-1)^s}{ n_1\cdot(n_1{+}n_2)\cdots(n_1{+}n_2{+}\ldots+n_s)}
\end{equation}
and related to the Stirling numbers of the second kind.
Writing out the first few terms and suppressing the functional argument~$\phi$ of $r_k$, 
the perturbative Nicolai map reads~\cite{LR1}
\begin{equation} \label{Texp}
\begin{aligned}
T_g\phi &\= \phi \ -\ g\,r_1 \phi \ -\ \sfrac12g^2\bigl(r_2-r_1^2\bigr)\phi\ -\
\sfrac16g^3\bigl(2r_3-r_1r_2-2r_2r_1+r_1^3\bigr)\phi \\
&\quad -\sfrac{1}{24}g^4\bigl(6r_4-2r_1r_3-3r_2r_2+r_1^2r_2-6r_3r_1
+2r_1r_2r_1+3r_2r_1^2-r_1^4\bigr)\phi \ +\ O(g^5)\, .
\end{aligned}
\end{equation}

For computing correlation functions \`a la (\ref{Nicdef}) we need the inverse map.
It possesses an analogous universal representation in terms of an anti-path-ordered exponential,
which gives rise to a different power series expansion,
\begin{equation}
T_g^{-1}\phi \= \!\sum_{\bf n} g^n\,d_{\bf n}\,r_{n_s}[\phi]\ldots r_{n_2}[\phi]\,r_{n_1}[\phi]\ \phi
\quad\textrm{with}\quad
d_{\bf n}  \= \frac{1}{n_s\cdot(n_s{+}n_{s-1})\cdots(n_s{+}n_{s-1}{+}\ldots{+}n_1)}
\end{equation}
whose first terms are
\begin{equation} \label{invTexp}
\begin{aligned}
T_g^{-1}\phi &\= \phi \ +\ g\,r_1 \phi \ +\ \sfrac12g^2\bigl(r_2+r_1^2\bigr)\phi\ +\ 
\sfrac16g^3\bigl(2r_3+2r_1r_2+r_2r_1+r_1^3\bigr)\phi \\
&\quad +\sfrac{1}{24}g^4\bigl(6r_4+6r_1r_3+3r_2r_2+3r_1^2r_2+2r_3r_1
+2r_1r_2r_1+r_2r_1^2+r_1^4\bigr)\phi \ +\ O(g^5)\, .
\end{aligned}
\end{equation}
Comparing this with~\eqref{Texp} it is clear how the two sets of coefficents are related.

This universal formula naturally generalizes to multiple couplings~\cite{LR3}.
Collecting a number $k$ of couplings in a formal vector,
\begin{equation}
\vec{g} \= \bigl( g^{(i)}\bigr) \= \bigl( g^{(1)}, g^{(2)}, \ldots, g^{(k)}\bigr)\ ,
\end{equation}
the individual flow equations read
\begin{equation}
\pa_{g^{(i)}}\bigl\<Y[\phi]\bigr\>_{\vec{g}} \=
\bigl\< \bigl(\pa_{g^{(i)}} + R^{(i)}_{\vec{g}}[\phi]\bigr)\,Y[\phi] \bigr\>_{\vec{g}}
\end{equation}
and define a formal vector 
$\vec{R}_{\vec{g}}:=\bigl(R^{(1)}_{\vec{g}}, R^{(2)}_{\vec{g}}, \ldots, R^{(k)}_{\vec{g}}\bigr)$
of flow operators (depending on all couplings). The universal formula now operates in a 
$k$-dimensional coupling space, thus the path-ordered exponential requires choosing a path
\begin{equation}
h:\ [0,1]\ \to\ \bigl\{\vec{g}\bigr\} \quad\und\quad t\ \mapsto\ \vec{h}(t) 
\quad\with \vec{h}(0)=\vec{0} \und \vec{h}(1)=\vec{g}\ .
\end{equation}
With these preliminaries, we obtain~\cite{LR3}
\begin{equation}
\begin{aligned}
T_{\vec{g}}\phi &\= {\cal P} \exp \Bigl\{-\!\!\int_0^{\vec{g}}\!\!\diff {\vec{g}}'\cdot\vec{R}_{\vec{g}'}[\phi]\Bigr\}\ \phi
\= {\cal P} \exp \Bigl\{-\!\!\int_0^1\!\!\diff t\ \vec{h}'(t)\cdot\vec{R}_{\vec{h}(t)}[\phi]\Bigr\}\ \phi \\
&\= \sum_{s=0}^\infty (-1)^s\!\!\int_0^1\!\!\diff t_s \ldots \!\int_0^{t_3}\!\!\!\!\diff t_2 \!\int_0^{t_2}\!\!\!\!\diff t_1\
\bigl[\vec{h}'(t_n)\cdot\vec{R}_{\vec{h}(t_s)}[\phi]\bigr] \ldots \bigl[\vec{h}'(t_1)\cdot\vec{R}_{\vec{h}(t_1)}[\phi]\bigr]\ \phi\ ,
\end{aligned}
\end{equation}
which upon Taylor-expanding $\vec{R}_{\vec{h}(t_i)}$ in powers of~$g^{(j)}$ produces a generalization of~\eqref{Tseries}.
We note that this map in general depends on the chosen path~$h$ in coupling space, despite
\begin{equation}
\bigl[ \pa_{g^{(i)}},\pa_{g^{(j)}} \bigr]\,\bigl\<Y\bigr\>_{\vec{g}} = 0 \quad\forall\,Y \quad\Rightarrow\quad
\bigl\< \pa_{g^{(i)}} R^{(j)}_{\vec{g}}[\phi] - \pa_{g^{(j)}} R^{(i)}_{\vec{g}}[\phi] 
+ \bigl[ R^{(i)}_{\vec{g}}[\phi],R^{(j)}_{\vec{g}}[\phi]\bigr] \bigr\>_{\vec{g}}= 0\ .
\end{equation}
This ``flatness condition'' in coupling space is only valid ``on the average''.
Hence, generically we have a functional family of Nicolai maps, yet all members yield the same correlation functions.
It is of course possible to consider paths that keep some couplings fixed, which then are ``spectating parameters''
(giving partial Nicolai maps) or to connect two finite-coupling values by flowing along some path leading from one to the other.

\begin{figure}
\begin{center}
\includegraphics[width=0.7\paperwidth]{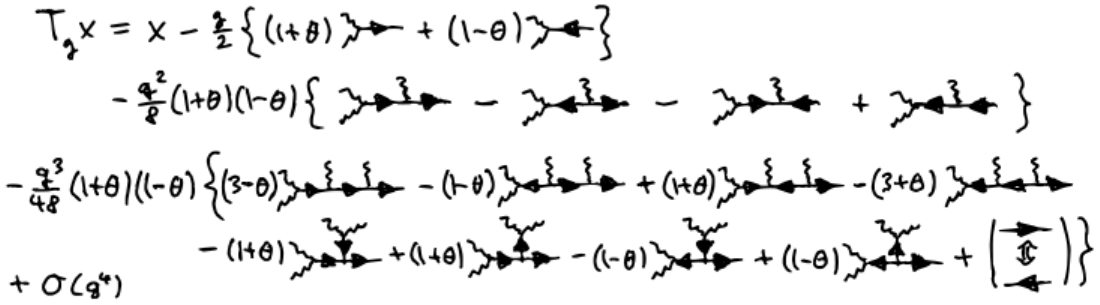}
\vspace{-6mm}
\end{center}
\caption{\label{fig5}Graphical Nicolai map expansion for supersymmetric quantum mechanics with theta term}
\end{figure}

In special cases, however, the map may be path-{\it in\/}dependent, and then it is unique!
In such a situation with more than one coupling, all powers in $g^{(i)}$ beyond the first cancel out,
and the power series truncates to a {\it linear\/} map,
\begin{equation}
T_{\vec{g}}\phi \= \phi \ -\ \vec{g}\cdot\vec{R}_{\vec{0}}[\phi]\ \phi\ .
\end{equation}
This is however not so for single-coupling flows, since these are path independent and in general non-polynomial.
Yet, even a single-coupling flow collapses to a linear map if only~\cite{LR3}
\begin{equation}
R_g[\phi]\,R_g[\phi]\,\phi\big|_{\textrm{no branch}} \= \bigl(\pa_g R_g[\phi]\bigr)\,\phi \qquad\Rightarrow\qquad
T_g\phi \= \phi - g\,r_1\,\phi\ ,
\end{equation}
where ``no branch'' means omitting all {\it branched\/} trees in the expansion.
This is not an empty condition but can happen for special fixed-coupling values. 
An example is supersymmetric quantum mechanics, say in one dimension for a bosonic trajectory~$x(t)$
and its Grassmann-valued fermionic partner~$\psi(t)$. With a cubic superpotential $W(x)=\sfrac12mx^2+\sfrac13gx^3$
and the inclusion of a total-derivative ``theta term'', the (on-shell) lagrangian reads ($\dot{x}=\pa_tx$)
\begin{equation}
L \= \sfrac12\dot{x}^2 -\sfrac12 m^2 x^2 - m g\,x^3 - \sfrac12g^2 x^4 +
\bar\psi [ \im\pa_t - m - 2g\,x]\psi + \im\,\theta\,(m\,x + g\,x^2)\,\dot{x}\ .
\end{equation}
Let us keep the mass~$m$ and the topological parameter~$\theta$ fixed and flow in the coupling~$g$ only.
Fourier-transforming from the time~($t$) to the frequency~$(\omega)$ domain,
the fermion propagator in $\omega$-space, depending on its direction, is denoted as
\begin{equation}
\sfrac{-1}{\omega+m} \=  -\!\!-\!\!\!\blacktriangleright\!\!\!-\!\!\!-  \qquad\und\qquad 
\sfrac{1}{\omega-m} \=   -\!\!-\!\!\!\blacktriangleleft\!\!\!-\!\!\!- \ .
\end{equation}
Depicting the $x$~insertions by wavy lines as usual, with energy conservation at each vertex
in frequency space, the diagrammatic representation of the Nicolai map up to order~$g^3$~\cite{LR3}
takes the form of Figure~\ref{fig5},
from which it is evident that the power series collapses for $\theta={\pm}1$.

\section{The case of gauge theories}

Suprsymmetric gauge theories present additional challenges for the Nicolai map.
Firstly, one has to deal with the gauge redundancy
necessitating a gauge fixing which in a Wess--Zumino gauge breaks supersymmetry and, secondly, 
the $g$-derivative of the supersymmetric action cannot easily be expressed as a supervariation
but also needs a BRST variation. Hence, Ward identities for BRST as well as broken supersymmetry
will be required~\cite{ANPP,ALMNPP}. The fermion matching condition for the map now equates 
its Jacobian with the product of the fermion and the Faddeev--Popov determinant, 
and the gauge-fixing condition should not be altered by the map.

We allow for a topological theta term (only in $D{=}4$) with $\theta'=\tfrac{g^2\theta}{8\pi^2}$,
use a local gauge-fixing functional $\cG$ to fix a gauge $\cG(A){=}0$ with a parameter~$\xi$
and include the corresponding ghost fields to formulate a BRST-invariant action
\begin{equation}
S_{\textrm{SUSY}}[A,\lambda,{\cal D},c,\bar c] \= \int\!\diff x\ \tr\bigl\{
-\tfrac14 F_{\mu\nu}F^{\mu\nu} +\tfrac{\theta'}{4} F_{\mu\nu}\widetilde{F}^{\mu\nu}
-\tfrac{\im}{2}\bar\lambda\slashed{D}\lambda + \tfrac12 {\cal D}^2
-\tfrac{1}{2\xi}\cG(A)^2 + \bar{c}\,\tfrac{\pa\cG}{\pa A_\mu}D_\mu c \bigr\}
\end{equation}
for $su(N)$-valued gluons~$A_\mu{=}A_\mu^A T^A$,
gluinos~$\lambda_\alpha{=}\lambda_\alpha^A T^A$,
an auxiliary~${\cal D}={\cal D}^A T^A$,
a ghost $c{=}c^A T^A$ and an antighost $\bar c{=}\bar c^A T^A$,
with $D_\mu=\pa_\mu+gA_\mu{\times}$ and
$F_{\mu\nu}{=}\pa_\mu A_\nu{-}\pa_\nu A_\mu{+}gA_\mu{\times}A_\nu{=}F_{\mu\nu}^AT^A$
and group generators subject to 
$[T^A,T^B]{\equiv}(T{\times}T)^{AB}{=}f^{ABC}T^C$ with $A,B,\ldots=1,2,\ldots,N^2{-}1$.
The trace refers to the color degrees of freedom.
We allow for various spacetime dimensionalities~$D$ by letting the fields live on~$\RR^{1,D-1}$
so that Lorentz indices $\mu,\nu,\ldots=0,1,\ldots,D{-}1$ and Majorana indices $\alpha=1,\ldots,r$,
where $r$ is the complex dimension of the corresponding Majorana representation,
i.e.~$\lambda^A\in\C^r$. It essentially grows exponentially with~$D$.

We know of two different Nicolai-map constructions, an off-shell one and an on-shell one.
In both cases, a gauge-fixing breaks supersymmetry.
The off-shell variant \cite{L1,MN,LR2} requires off-shell supersymmetry, thus works only 
(for a finite number of auxiliary fields) in four or less spacetime dimensions. 
It parallels the construction for chiral multiplets, employing gauge superfields, 
and admits any choice of gauge fixing.
The on-shell version \cite{FL,ALMNPP} works in higher dimensions but relies on an ansatz 
for the coupling flow operator, which however does its job only partially,
\begin{equation}
\pa_g\bigl\<Y[A]\bigr\>_g \= \bigl\< \bigl(\pa_g+R_g[A]+Z_g[A]\bigr)\,Y[A]\bigr\>_g\ .
\end{equation}
Here, a {\it multiplicative\/} contribution~$Z_g$ destroys the derivation property of~$R_g$ and hence
the distributivity of~$T_g$, which is not acceptable.
A somewhat lengthy computation reveals, however, that in the Landau gauge,
$\cG{=}\pa^\mu A_\mu$ with $\xi{\to}\infty$, the obstacle may be overcome because
\begin{equation}
Z_g=0 \qquad\textrm{if and only if}\quad 
r=2(D{-}2) \quad\Leftrightarrow\quad  D=3,4,6,10\ .
\end{equation}
Amazingly, these are precisely the ``critial spacetime dimensions''
which admit super Yang--Mills theory to exist~\cite{BSS},
demonstrating that the Nicolai map knows about them~\cite{ANPP}!

In both versions, the most simple form of the coupling flow operator and thus the Nicolai map
arises in the Landau gauge on the gauge hypersurface $\pa^\mu A_\mu=0$,
\begin{equation} \label{gaugeflow}
\overleftarrow{R}_g[A] \= -\tfrac{1}{2r} \smallint\!\!\smallint\!\!\smallint\tr\
\overleftarrow{\tfrac{\delta}{\delta A_\mu}}\,P_\mu^{\ \nu}\,\bigl\{
\gamma_\nu\,\bcontraction{}{\bar\lambda}{\ }{\lambda}
\bar\lambda\ \lambda\,\gamma^{\rho\sigma} \bigl[\unity{+}\im\theta'\gamma^5\bigr] 
A_\rho{\times}A_\sigma \bigr\}_{\alpha\alpha}
\end{equation}
with the non-Abelian transversal projector~\cite{ALMNPP}
\begin{equation}
\begin{aligned} \label{proj}
P_\mu^{\ \nu} &\=
\delta_\mu^{\ \nu}\unity\ -\ D_\mu \bcontraction{}{c}{\ }{{\bar c}}c\ \bar c\,\pa^\nu
\qquad\Rightarrow\qquad
\pa^\mu\,P_\mu^{\ \nu} = 0 = P_\mu^{\ \nu} D_\nu \\
&\= \delta_\mu^{\ \nu}\unity\ -\ \pa_\mu \Box^{-1} \pa^\nu\ +\
g(A_\mu-\pa_\mu\Box^{-1} A{\cdot}\pa)(\pa{\cdot}D)^{-1}\pa^\nu\\[4pt]
&\ =:\ (P^{\textrm{inv}})_\mu^{\ \nu} + (P^{\textrm{lgt}})_\mu^{\ \nu} + (P^{\textrm{gh}})_\mu^{\ \nu}
\end{aligned}
\end{equation}
forcing the coupling flow onto the gauge surface: $R_g\cG\sim\cG$.
We have reversed the direction of the derivatives since acting towards the left is more
convenient for the graphical representation.
Due to the identity $D^\nu\big\{\ldots\bigr\}_\nu=0$, the three pieces of $P_\mu^{\ \nu}$ 
named above are of order $g^0$, $g^1$ and $g^2$, respectively.

The upshot is that our explicit construction formula~(\ref{universal}) carries over to gauge theory, 
for $D{\le}4$ in any gauge and for $D{=}6$ and~$10$ in the Landau gauge~\cite{LR2},
\begin{equation} \label{universalA}
T_g A\= {\cal P} \exp \Bigl\{-\!\int_0^g\!\!\diff h\ R_h[A]\Bigr\}\ A
\= \sum_{\bf n} g^n\,c_{\bf n}\,r_{n_s}[A]\ldots r_{n_2}[A]\,r_{n_1}[A]\ A\ .
\end{equation}
Here, we have expanded the coupling flow operator into homogeneous pieces,
\begin{equation}
R_g[A] \= r_1[A]\ +\ g\,r_2[A]\ +\ g^2 r_3[A]\ +\ldots
\qquad\textrm{with}\qquad \smallint A\tfrac{\delta}{\delta A}\,r_k[A] = k\,r_k[A]\ ,
\end{equation}
where the Taylor coefficients~$r_k$ decompose according to~\eqref{proj},
\begin{equation}
r_k \= r_k^{\textrm{inv}} + r_k^{\textrm{lgt}} + r_k^{\textrm{gh}}
\quad\with r_1^{\textrm{lgt}}{+}r_1^{\textrm{gh}}=0=r_2^{\textrm{gh}}\ .
\end{equation}

\begin{figure}
\begin{center}
\includegraphics[width=0.7\paperwidth]{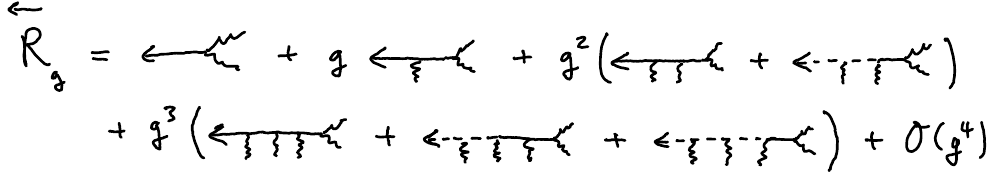}
\vspace{-6mm}
\end{center}
\caption{\label{fig6}Yang--Mills coupling flow operator in Landau gauge}
\end{figure}

\begin{figure}
\begin{center}
\includegraphics[width=0.7\paperwidth]{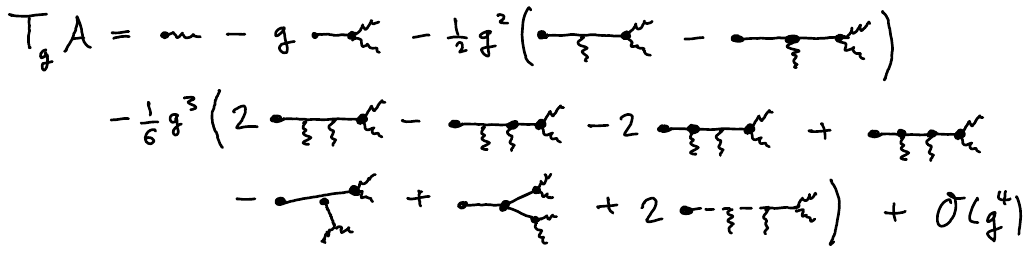}
\vspace{-6mm}
\end{center}
\caption{\label{fig7}Yang--Mills Nicolai map in Landau gauge}
\end{figure}

The diagrammatics in the Landau gauge looks fairly simple~\cite{FL,ALMNPP}.
With the solid line representing the free fermion propagator $(\im\slashed\pa)^{-1}$
and the dashed line standing for the free ghost propagator $\Box^{-1}$, 
we obtain for $\overleftarrow{R}_g[A]$ a linear tree expansion~\cite{ALMNPP} as in Figure~\ref{fig6}.
Iterating this functional differential operator in the universal formula~(\ref{universalA}) produces
(with rules analogous to the scalar case) the graphical representation 
of the Nicolai map~\cite{ALMNPP} is given in Figure~\ref{fig7}.
where the color structure follows the graphical one,
and we have suppressed the Lorentz and spinor indices.
In fact, performing the spin traces creates various contractions
of Lorentz indices on the gauge-field legs and on the propagators due to
$(\im\slashed{\pa})^{-1}\!=\im\gamma^\mu\pa_\mu\Box^{-1}$,
so that the number of terms at $O(g^n)$ grows rapidly with~$n$.
Nevertheless, the expansion is algorithmic and may be implemented on a computer.

For $\theta{=}0$ the map has been evaluated to order~$g^3$ in 2020~\cite{ALMNPP} and
pushed to order~$g^4$ one year later~\cite{MN}. 
Let us go beyond and allow for the topological term in $D{=}4$. Abbreviating 
\begin{equation}
\Box^{-1} = C \quad\und\quad \pa_\alpha\Box^{-1}=C_\alpha \qquad \textrm{etc.}
\end{equation}
we obtain (now $r{=}4$)
\begin{align} \label{Tgauge}
T_gA_\mu &\= A_\mu \ -\ g\,r_1 A_\mu \ -\ \sfrac12g^2\bigl(r_2-r_1^2\bigr)A_\mu\ -\ 
\sfrac16g^3\bigl(2r_3-2r_2r_1-r_1r_2+r_1^3\bigr)A_\mu\ +\ O(g^4) \\[4pt] \nonumber
&\= A_\mu \ -\ g\bigl( C^\lambda A_\mu{\times}A_\lambda 
-\tfrac12\theta'\epsilon_{\mu\nu\rho\lambda}C^\nu A^\rho{\times}A^\lambda\bigr)
-\ \tfrac32 g^2[1{+}{\theta'}^2]\,
C^\rho A^\lambda{\times}C_{[\rho}A_\mu{\times}A_{\lambda]}\ +\ O(g^3)\ .
\end{align}
Reminiscent of the map collapse for supersymmetric quantum mechanics in the previous section,
we anticipate simplifications for the magical values $\theta'={\pm}\im$, where at least
the second-order contribution cancels out! And indeed, for these two choices the square
bracket in~\eqref{gaugeflow} becomes $\unity\mp\gamma^5$, which is (twice) the chiral spin projector.
In this special case, the Fierz identity 
\begin{equation}
\tr\bigl( Y\,\tfrac12[\gamma^\mu(\unity{+}\gamma^5]\bigr)\cdot
\tr\bigl( \tfrac12[\gamma_\mu(\unity{-}\gamma^5]\,Z\bigr)
\= -2\,\tr\bigl( Y\,\tfrac12[\unity{-}\gamma^5]\,Z\,\tfrac12[\unity{+}\gamma^5]\bigr)
\qquad\forall\ Y, Z
\end{equation}
allows us to fuse two spin traces along the tree, via~\cite{LR4}
\begin{equation}
r_{k-1}^{\textrm{inv}}\;r_1\;A\= (r_{k}^{\textrm{inv}}+r_{k}^{\textrm{lgt}})\;A
\qquad\textrm{for}\quad \theta'=\pm\im\quad\textrm{and}\quad k\ \geq\ 2\ .
\end{equation}
Inserting this into~\eqref{Tgauge}, the $O(g^2)$ term vanishes because $r_1^2A=r_2A$, 
and many terms at higher orders cancel systematically, for example the first branched tree
$r_1^3A$ at $O(g^3)$. Some `gh' and `lgt' contributions survive, however, 
and to third order in~$g$ at $\theta'{=}{-}\im$ we find
\begin{equation}
\begin{aligned}
T_gA_\mu &\= A_\mu \ -\ g\;r_1A_\mu \ -\ \sfrac13g^3\bigl(r_3^{\mathrm{gh}}-r_2^{\mathrm{lgt}}r_1\bigr)A_\mu\ +\ O(g^4)\\[4pt]
&\= A_\mu\ +\ \tfrac18 g\;\tr\bigl\{\gamma_{\mu\alpha}\gamma^{\rho\lambda}\bigl[\unity{+}\gamma^5\bigr]\bigr\}\,
C^\alpha A_\rho{\times}A_\lambda \\
&\qquad\quad +\ \tfrac1{24}g^3\, [A_\mu-C_\mu A{\cdot}\partial]\,C\,A^\nu\;
\tr\bigl\{\gamma_\nu\gamma_\beta\gamma^{\rho\lambda}\bigl[\unity{+}\gamma^5\bigr]\bigr\}\,
C^\beta A_\rho{\times}A_\lambda \\
&\qquad\quad -\ \tfrac1{96}g^3\,\tr\bigl\{ \gamma_{\mu\alpha}\gamma^{\nu\beta}\bigl[\unity{+}\gamma^5\bigr]\bigr\} \;
\tr\bigl\{ \gamma_{\sigma\gamma}\gamma^{\rho\lambda}\bigl[\unity{+}\gamma^5\bigr]\bigr\}\,
C^\alpha A_\nu C_\beta A^\sigma C^\gamma A_\rho{\times}A_\lambda\ +\ \mathrlap{O(g^4)\ .}
\end{aligned}
\end{equation}
Performing the spin traces, one arrives at a ``chiral Nicolai map''~\cite{LR4}:
\begin{equation}
\begin{aligned}
T_gA_\mu \= &A_\mu\ 
-\ g\ \bigl\{ C_\lambda A_\mu{\times} A^\lambda
+ \sfrac{\im}{2} \epsilon_{\mu\alpha\rho\lambda} C^\alpha A^\rho{\times} A^\lambda \bigr\}\\            
-\, &\sfrac{g^3}{3}\,[A_\mu-C_\mu A\cdot\partial]\, C A_\rho C_\lambda A^\rho{\times} A^\lambda\\
+\, &\sfrac{2g^3}{3}\,C^\alpha A_{[\mu} C_{\alpha]} A_\rho C_\lambda A^\rho{\times} A^\lambda \,
+\, 4 g^3\,C^\nu A^\alpha C^\beta A_{[\mu} C_\nu A_\alpha {\times} A_{\beta]}\\
-\,&\sfrac{\im g^3}{6}\epsilon_{\nu\sigma\rho\lambda} [A_\mu-C_\mu A\cdot\partial] C A^\nu C^\sigma A^\rho{\times} A^\lambda\,
+\, \sfrac{\im g^3}{3}\epsilon_{\nu\sigma\rho\lambda} C^\alpha A_{[\mu} C_{\alpha]} A^\nu C^\sigma A^\rho{\times} A^\lambda \\
-\, &\sfrac{\im g^3}{3}\epsilon_{\mu\nu\alpha\beta}C^\alpha A^\nu C^\beta A_\rho C_\lambda A^\rho{\times} A^\lambda \
+\ O(g^4)\ .
\end{aligned}
\end{equation}
As in the quantum mechanical example, the imaginary terms are of no concern: 
although the chiral map is no longer real, imaginary contributions to real correlation functions always cancel out
because only even powers of~$\theta'$ survive.
This chiral version of the map was pushed to $O(g^4)$ and tested to $O(g^3)$~\cite{LR4}. 

\begin{wraptable}{r}{6.2cm}
\vspace{-0.5cm}
\caption{\label{tab1}Number of terms in $T_gA$}
{\small
\hspace{0.3cm}
\begin{tabular}{ l  c c c c }
\br
order $n$ in $g$ & \ 1\  & \ 2 & \ 3 & 4 \\
\mr
non-chiral map & \ 1\ & \ 3 & \ 34 & 380 \\
chiral map & \ 2\ & \ 0 & \ 21 & 224 \\
\br
\end{tabular}
}
\vspace{-0.6cm}
\end{wraptable}
Taking into account all the antisymmetrizations of indices while respecting the symmetries of the various topologies,
we count the number of terms in the first four orders for the non-chiral map~\cite{MN} versus the chiral map~\cite{LR4}.
There does not seem to be a huge difference between the two formulations,
but the epsilon symbol generated in the chiral map allows one to combine the antisymmetrization of many terms.

\section{Summary}
\begin{itemize}
\item Globally supersymmetric theories can be investigated from a novel perspective using Nicolai maps
\item Efficient computation of amplitudes via alternative graphical ``Nicolai rules''~\cite{DL2,NP,M}
\item Convergent power series for the functional transformation to the free theory~\cite{L3}
\item Works also for super Yang--Mills in 3, 4, 6 and 10 spacetime dimensions
\item Ambguities: path dependence in coupling space, $\theta$ angle, R-symmetry~\cite{R}
\item Special $\theta$ values project onto single helicity and may collapse the map: hope for $D{=}4$ ${\cal N}{=}\,4$ SYM?
\item Quantization of the maximally supersymmetric membrane via the supermatrix model?~\cite{LN}
\item Perspectives: SYM amplitudes, integrability, supersymmetric sigma models, supergravity?
\item Question: for which geometries can the Riemann tensor be a sum of squares?
\end{itemize}


\section*{References}
\begin{thebibliography}{9}

\bibitem{Nic1}
Nicolai H 1980
On a new characterization of scalar supersymmetric theories,
\href{https://dx.doi.org/10.1016/0370-2693(80)90138-0}
{{\it Phys.\ Lett.\ B} {\bf 89} 341}

\bibitem{Nic2}
Nicolai H 1980
Supersymmetry and functional integration measures,
\href{https://dx.doi.org/10.1016/0550-3213(80)90460-5}
{{\it Nucl.\ Phys.\ B} {\bf176} 419}

\bibitem{Nic3}
Nicolai H 1984
Supersymmetric functional integration measures,\\
{\sl NATO Advanced Study Institute on Supersymmetry (Bonn)},
\href{https://cds.cern.ch/record/155731?ln=en}
{ed K~Dietz et al (Plenum Press) p 393}

\bibitem{L1}
Lechtenfeld O 1984
Construction of the Nicolai mapping in supersymmetric field theories,\\
{\it Ph.D.\ thesis Bonn University},
\href{https://lib-extopc.kek.jp/preprints/PDF/2000/0030/0030157.pdf}
{internal report {\it BONN-IR-84-42}, ISSN-0172-8741}

\bibitem{DL1}
Dietz K and Lechtenfeld O 1985
Nicolai maps and stochastic observables from a coupling constant flow,\\
\href{https://dx.doi.org/10.1016/0550-3213(85)90132-4}
{{\it Nucl.\ Phys.\ B} {\bf 255} 149}

\bibitem{DL2}
Dietz K and Lechtenfeld O 1985
Ghost-free quantisation of non-Abelian gauge theories
via the Nicolai transformation of their supersymmetric extensions,
\href{https://dx.doi.org/10.1016/0550-3213(85)90642-X}
{{\it Nucl.\ Phys.\ B} {\bf 259} 397}

\bibitem{L2}
Lechtenfeld O 1986
Stochastic variables in ten dimensions?,
\href{https://doi.org/10.1016/0550-3213(86)90531-6}
{{\it Nucl.\ Phys.\ B} {\bf 274} 633}

\bibitem{FL}
Flume F and Lechtenfeld O 1984
On the stochastic structure of globally supersymmetric field theories,\\
\href{https://doi.org/10.1016/0370-2693(84)90459-3}
{{\it Phys.\ Lett.\ B} {\bf 135} 91}

\bibitem{LR1}
Lechtenfeld O and Rupprecht M 2021
Universal form of the Nicolai map,
\href{https://doi.org/10.1103/PhysRevD.104.L021701}
{{\it Phys.\ Rev.\ D} {\bf 104} L021701}
(\href{https://arxiv.org/abs/2104.00012}{{\it Preprint} 2104.00012})

\bibitem{LR3}
Lechtenfeld O and Rupprecht M 2022
Is the Nicolai map unique?,
\href{https://doi.org/10.1007/JHEP09(2022)139}
{{\it J.\ High Energ.\ Phys.} 139} 
(\href{https://arxiv.org/abs/2207.09471}{{\it Preprint} 2207.09471})

\bibitem{ANPP}
Ananth S, Nicolai H, Pandey C and Pant S 2020
Supersymmetric Yang--Mills theories: not quite the usual perspective,\\
\href{https://doi.org/10.1088/1751-8121/ab7b9d}
{{\it J.\ Phys.\ A: Math.\ Theor.} {\bf 53} 174001}
(\href{https://arxiv.org/abs/2001.02768}{{\it Preprint} 2001.02768})

\bibitem{ALMNPP}
Ananth S, Lechtenfeld O, Malcha H, Nicolai H, Pandey C and Pant S 2020\\
Perturbative linearization of supersymmetric Yang--Mills theory,
\href{https://dx.doi.org/10.1007/JHEP10(2020)199}
{{\it J.\ High Energ.\ Phys.} 199}
(\href{https://arxiv.org/abs/2005.12324}{{\it Preprint} 2005.12324})

\bibitem{MN}
Malcha H and Nicolai H 2021
Perturbative linearization of super-Yang--Mills theories in general gauges,\\
\href{https://doi.org/10.1007/JHEP06(2021)001}
{{\it J.\ High Energ.\ Phys.} 001}
(\href{https://arxiv.org/abs/2104.06017}{{\it Preprint} 2104.06017})

\bibitem{LR2}
Lechtenfeld O and Rupprecht M 2021
Construction method for the Nicolai map in supersymmetric Yang--Mills theories,\\
\href{https://doi.org/10.1016/j.physletb.2021.136413}
{{\it Phys.\ Lett.\ B} {\bf 819} 136413}
(\href{https://arxiv.org/abs/2104.09654}{{\it Preprint} 2104.09654})

\bibitem{BSS}
Brink L, Schwarz JH and Scherk J 1977
Supersymmetric Yang--Mills theories,
\href{https://doi.org/10.1016/0550-3213(77)90328-5}
{{\it Nucl.\ Phys.\ B} {\bf 121} 77}

\bibitem{LR4}
Lechtenfeld O and Rupprecht M 2023
An improved Nicolai map for super Yang--Mills theory,\\
\href{https://doi.org/10.1016/j.physletb.2023.137681}
{{\it Phys.\ Lett.\ B} {\bf 838} 137681}
(\href{https://arxiv.org/abs/2211.07660}{{\it Preprint} 2211.07660})

\bibitem{NP}
Nicolai H and Plefka J 2020
${\cal N}{=}\,4$ super-Yang--Mills correlators without anticommuting variables,\\
\href{https://doi.org/10.1103/PhysRevD.101.125013}
{{\it Phys.\ Rev.\ D}  {\bf 101} 125013}
(\href{https://arxiv.org/abs/2003.14325}{{\it Preprint} 2003.14325})

\bibitem{M}
Malcha M 2022
Two loop ghost free quantisation of Wilson loops in ${\cal N}{=}\,4$ supersymmetric Yang--Mills,\\
\href{https://doi.org/10.1016/j.physletb.2022.137377}
{{\it Phys.\ Lett.\ B} {\bf 833} 137377}
(\href{https://arxiv.org/abs/2206.02919}{{\it Preprint} 2206.02919})

\bibitem{L3}
Lechtenfeld O 2022
Supersymmetric large-order perturbation with the Nicolai map,\\
\href{https://doi.org/10.1016/j.physletb.2022.137507}
{{\it Phys.\ Lett.\ B} {\bf 835} 137507}
(\href{https://arxiv.org/abs/2208.06420}{{\it Preprint} 2208.06420})

\bibitem{R}
Rupprecht M 2022
The coupling flow of ${\cal N}{=}\,4$ super Yang--Mills theory,
\href{https://doi.org/10.1007/JHEP04(2022)004}
{{\it J.\ High Energ.\ Phys.} 004}
(\href{https://arxiv.org/abs/2111.13223}{{\it Preprint} 2111.13223})

\bibitem{LN}
Lechtenfeld O and Nicolai H 2022
A perturbative expansion scheme for supermembrane and matrix theory,\\
\href{https://doi.org/10.1007/JHEP02(2022)114}
{{\it J.\ High Energ.\ Phys.} 114}
(\href{https://arxiv.org/abs/2109.00346}{{\it Preprint} 2109.00346})

\end{thebibliography}
\end{document}